\documentclass{elsart}
\usepackage{epsfig}

\begin{document}

\begin{frontmatter}

\title{SUE: A Special Purpose Computer for Spin Glass Models}

\author[Zaragoza]{A. Cruz},
\author[Zaragoza,Otro]{J. Pech},
\author[Zaragoza]{A. Taranc\'on}, 
\author[SIC]{P. T\'ellez}, 
\author[Zaragoza]{C. L. Ullod} and 
\author[Zaragoza]{C. Ungil}.

\address[Zaragoza]{Departamento de F\'{\i}sica Te\'orica,
        Facultad de Ciencias, \\
        Universidad de Zaragoza, 50009 Zaragoza, Spain\\
\small e-mail: \tt tarancon@sol.unizar.es}

\address[Otro]{Institute of Physics, Academy of Sciences, \\
        180 40  Prague, Czech Republic}
\address[SIC]{Servicio de Instrumentaci\'on Cient\'{\i}fica
        Facultad de Ciencias, \\
        Universidad de Zaragoza, 50009 Zaragoza, Spain\\}


\begin{abstract}
The use of last generation Programmable Electronic Components 
makes possible the construction of very powerful and competitive 
special purpose computers. 
We have designed, constructed and tested a three-dimensional Spin Glass model
dedicated machine, which consists of 12 identical boards.
Each single board can simulate 8 different systems, updating all the 
systems at every clock cycle. 
The update speed of the whole machine is 217ps/spin with 48~MHz clock
frequency.
A device devoted to fast random number generation has been developed 
and included in every board.
The on-board reprogrammability permits us to change easily the 
lattice size, or even the update algorithm or the action.
We present here a detailed description of the machine and the 
first runs using the Heat Bath algorithm. 
\end{abstract}

\end{frontmatter}

\vfill
DFTUZ/2000/02 \hfill{cond-mat/0004080}
\vfill

\section{Introduction}

Two approaches have become popular in the field of computer design 
for scientific calculations: special or general purpose computers.
Lattice Monte Carlo in Quantum Field Theory and Statistical Mechanics 
requires large computational power in relatively general 
purpose computers and the processing can often be parallelized. 
Various groups have developed their own parallel machines for those 
simulations~\cite{APE}~\cite{CHRIST}~\cite{RTN}~\cite{HOOGLAND}.
Those general purpose computers require continuous technological 
upgrading and investment to obtain competitive results.
On the other hand, special purpose computers can approach very specific
problems, achieving better performance than general computers.

The emergence in the market of Complex Programmable Logic Devices 
(CPLD) makes it possible to design dedicated machines with 
low cost and high performance. 
In this paper we describe a CPLD-based machine, dedicated to
three-dimensional spin glass models with variables belonging to $Z_2$
and couplings to first neighbours, and report on the reliability tests
which have been carried out.

Our machine is called SUE, for Spin Updating Engine, because its
task is to generate sets of updated spin configurations in
the Monte Carlo simulation.
In a previous work~\cite{SUE2D}, we presented the prototype for the 
two-dimensional model, and introduced the first ideas about the final 
version.
After checking that the 2d version worked properly, we have
designed, constructed and tested the 3d version, which differs 
in some aspects from the 2d version as we will see below. 
The performance of the 3d machine is improved due to the fact that it can 
run more than a single model: the lattice size or the action of the 
physical model can be easily changed using the on-board reprogrammability 
of the CPLDs. A device devoted to generate a 32-bit random number  
has been developed and included in every SUE board. This device 
(described 
below) enables SUE to operate with both canonical and microcanonical 
algorithms.

At present, spin glass models~\cite{SG} are a progressing area of 
Statistical Mechanics. They are related to neural networks, spin models, 
some High $T_c$ superconductivity models, etc.
There is large activity in the 3d models because of the uncertainty 
in the vacuum structure at low temperature. Monte Carlo simulations of
spin glass systems have been used to study the phase transition, 
the ultrametric structure and the dynamics out of equilibrium~\cite{TEMP}.
Only sizes up to $L=16$ have been simulated~\cite{MARI}\cite{BERG}\cite{YOUNG},
due to the slow dynamics of the systems and the strong slowing-down
as the size grows. 
Yet, those simulations requiring very simple calculations, they are easily 
implementable in a dedicated machine. That way the computational 
power needed to obtain results in larger lattices is obtained.

A standard way of studying spin glasses is the use of independent lattices
with the same quenched couplings, called {\it replicas}.
The overlap between two {\it replicas} acts as the order parameter 
in that model.
A great improvement on the usual Monte Carlo scheme is the parallel 
tempering method~\cite{TEMP}. The basic idea is to move in temperature 
space: the system changes its temperature, goes up to the paramagnetic 
phase and  eventually goes back to lower temperature. With high 
probability in its motion through temperature the system will visit new 
local minima.
That scheme has been implemented in SUE: {\it Replicas} at different 
temperatures are simulated, and systems running at adjacent levels can 
be swapped according to an appropriate probability distribution.

An essential tool for the analysis of results is Finite Size
Scaling~\cite{FISCHER}, which requires the use of different volumes.
In that sense, SUE is capable of working with different lattice sizes by
reprogramming its CPLD devices.

The main differences with respect to the 2d prototype presented in~\cite{SUE2D} are:
\begin{itemize}
\item{} Larger and faster devices.
\item{} Multi-Layer instead of Double-Layer Printed Circuits.
\item{} On-board reprogrammability.
\item{} Dedicated device for 32-bit random number generation.
\item{} Demon and Heat Bath algorithm support.
\item{} Parallel Tempering implementation.
\item{} Driver and Software development for easy (transparent) use.
\end{itemize}

At present we have built 12 boards and tested them using the 
Demon and Heat Bath algorithms in different lattice sizes. 
Each board simulates 8 lattices, updating 8 spins every $20.8ns$ cycle.
The update speed of a single board is therefore $2.6ns/spin$.
The cost of each board is 2400 Euros (500 for PCB and mounting, and
1900 for components.)

The summary of this paper is as follows: We start introducing
the physical model in the next section.
In section~\ref{OPERATION}
We describe the electronic architecture of SUE, design considerations
and software support. The development process is outlined in 
section~\ref{DEVELOPMENT}. Last section is devoted to 
discuss the performance.

\section{The Physical Model}

We want to simulate the 3D Edwards-Anderson model
with first neighbour couplings (see~\cite{SG} for a detailed
description of the model). The action of this model is given by 
\begin{equation}
E=\sum_{i,j} \sigma_i \sigma_j J_{ij}, 
\label{action}
\end{equation}
where the value of the Ising spins $\sigma$ can be $1$ or $-1$,
and the couplings $J_{ij}$ are random variables taking the values
$\pm1$ with equal probability. For a fixed set of couplings 
$\{J_{ij}\}$ the partition function is
\begin{equation}
Z(\beta,\{J_{ij}\})=\sum_{\{\sigma\}} \exp{\beta E(\{J_{ij}\},\{\sigma\}}).
\label{partition}
\end{equation}

We study the existence of phase transitions using as order parameter the
overlap between two independent systems ({\it replicas}) with the same set 
of couplings $J_{ij}$. We should finally average over different realizations 
of the disorder ($J_{ij}$) to obtain physical results about the system.

To calculate (\ref{partition}) we must sum over $2^V$ possible 
configurations, where $V$ is the volume of the lattice, which is a 
very large number for any computer. The standard way to compute the 
partition function is to run an algorithm that selects only a representative 
set of configurations. There are different appropiate algorithms, see for 
instance the chapter by Sokal in~\cite{QFC}. 
For pure spin systems 
(all $J_{ij}$ equal to $1$) some {\it cluster} algorithms are very efficient, 
but for a general spin glass model only local algorithms achieve 
good efficiency. 

Typically, we must run the algorithm and generate 
millions of different representative configurations in order to obtain 
accurate results. The {\it autocorrelation time} $\tau$ is a measure 
of the correlation between configurations: a run of length $n$ provides 
only $\sim n/\tau$ effectively independent samples. Near the critical 
point, $\tau$ diverges as $\tau\sim L^z$, where $L$ is the size of the system 
and $z$, the {\it dynamical critical exponent}, has been found to be around 
$6$ (while in the pure Ising model it is close to $2$).
This strong slowing-down, due to the existence of many pure 
and metastable states, and the absence of non local algorithms, makes this 
problem really hard from a computational point of view.

Two different updating algorithms have been implemented in our design,
one microcanonical (Demon) and one canonical (Heat Bath).

The Demon algorithm~\cite{CREUTZ}\cite{CluJJ}
keeps the sum of the lattice energy and a demon energy constant. 
In order to generate the representative set of samples, 
we start from a spin configuration with an action $S$ and a demon energy 
equal to zero. 
Now we use the algorithm to change the spins to generate new configurations 
(one for every $V$ updates). The update of a spin is as follows: if the flip 
lowers the spin energy, the demon takes that energy and the flip is accepted. 
On the other hand, if the flip increases the spin energy, the change is only made 
if the demon energy is sufficient to  transfer that energy to the system.
The conservation of the total energy (lattice plus demon), has been useful
in the programming/test stage, allowing fast tests of proper function. 

In the Heat Bath algorithm~\cite{QFC}, the new spin value for each site $\sigma_i$ 
is independent from the old one, and its probability distribution is that 
of a single Ising spin $\sigma_i$ in the effective magnetic field produced 
by the fixed neighbouring spins $\sigma_j$:
\begin{equation}
P(\sigma_i\mid\{{\sigma_j}\}_{j\neq i})=
\frac{\exp{(\beta\sigma_i\sum_j{J_{ij}\sigma_j})}}
{
{\exp{(\beta\sum_j{J_{ij}\sigma_j})}+\exp{(-\beta\sum_j{J_{ij}\sigma_j})}}}.
\label{probdist}
\end{equation}

The drawback of this canonical algorithm is the necessity of a random 
number to decide the acceptance of the new spin.
 
The algorithms being local, to update a spin only the nearest neighbours
are needed. Because of simplicity in the electronic design, 
we use helicoidal boundary conditions.
Let us consider a lattice of side $L$ and volume $V=L^3$ with sites 
labelled in the standard way: the site $[x,y,z]$ (with 
$0 {\leq} x,y,z {\leq} L-1$) gets the index $n=x+y{\times}L+z{\times}L^2$.
We will call $x_+$ ($y_+$, $z_+$) the neighbour in the positive 
direction along the $x$ ($y$, $z$) axis. With our helicoidal boundary 
conditions the neighbours of the site $n$ are simply:
\begin{equation}
\begin{array}{ccc}
x_+ & = & (n+1) \bmod V\\
y_+ & = & (n+L) \bmod V\\
z_+ & = & (n+L^2) \bmod V\\
\end{array}
\label{hbc}
\end{equation}

We define in an analogous manner the remaining neighbours $x_-$, $y_-$ and $z_-$.

\begin{figure}[htb]
\centering
\vglue 0.3cm
\epsfig{figure=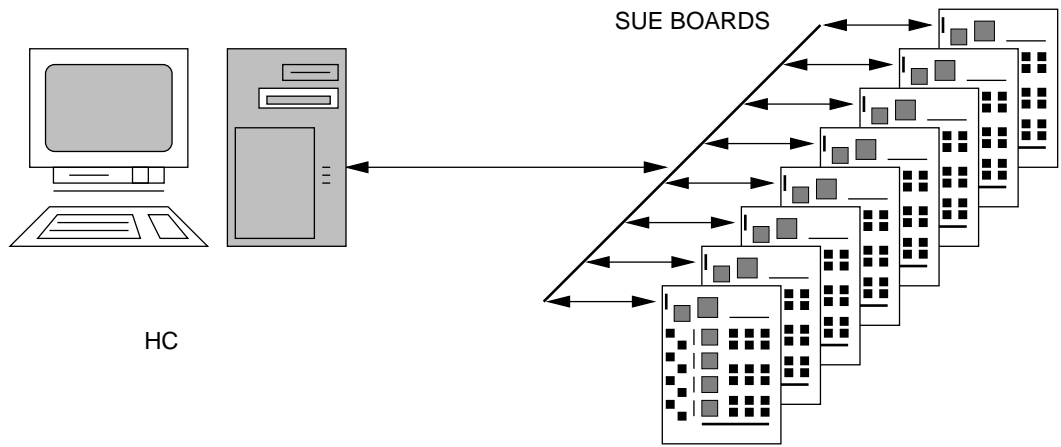,angle=0,width=140mm}
\caption{Schematic view of the full d=3 machine.}
\label{suefig}
\end{figure}          

\section{Operation and General Structure of SUE \label{OPERATION}}

The SUE machine is connected 
to a Host Computer (HC) running under Linux. SUE performs the update of the 
configurations, but the measurements and analysis are made by the HC.
SUE is set up with initial spin configurations, couplings and several
simulation parameters. Then SUE is started and simulation begins.   
After a certain number of iterations, SUE is stopped to download
the configuration to the HC and SUE keeps the updating process.
In this sense, SUE and the HC work in parallel: while SUE is updating the 
system the HC processes the previously read configurations.

Fig.~\ref{suefig} shows a simple diagram of the whole machine 
which consists
of the HC and $n$ SUE boards (the figure is for $n=8$, but the final system 
consists of 12 processing modules). They are connected to the HC 
through a PCI Data Acquisition Card. 
Every processing module contains the hardware to store 
and update eight lattices in parallel. Note that there are two degrees of 
parallelism: inside the processing module and between the modules.

Every clock cycle, the random number generator device included 
in each board provides a pseudo-random
number which is shared for the update of the eight lattices, so 
the {\it replicas} (systems with
the same couplings $J_{ij}$) must be simulated in different boards.
We can then think of each pair of boards as a unit, allowing us to
simulate eight pairs of {\it replicas} (corresponding to eight realizations 
of disorder $J_{ij}$). Periodically, the configurations are read
and the relevant measurements carried out and stored. 

Parallel tempering requires the simulation of pairs of {\it replicas} with
the same couplings at different values of $\beta$.
With 12 boards we could then simulate {\it replicas} corresponding 
to eight sets 
of couplings at 6 values of $\beta$ at once.
Parallel tempering requires more temperature levels, so each time
the configuration is read the $\beta$ is changed (the corresponding
probability table is loaded), and the configuration to be updated is
loaded onto the board (in the meantime it was stored in the HC).
Different temperatures are then sequentially simulated.

The HC controls this mechanism, and is responsible for deciding whether
the configurations being simulated at adjacent temperatures are 
interchanged. Given the configurations $X$ at temperature $\beta$
and $X'$ at temperature $\beta$', we compute
\begin{equation}
\Delta=({\beta}'-{\beta})(E(X)-E(X'))
\label{parallel}
\end{equation}
and use a Metropolis like test: if $\Delta < 0$ we accept the change,
otherwise we swap the configurations with probability $\exp{(-\Delta)}$. 

By processing 8 spins in parallel on 12 modules (96 spins in total) within 
one clock cycle (clock period of 48~MHz), we obtain an update speed of 
217 {\it ps/spin}. 
The time spent reading, writing and processing (Meassurement and Paralell Tempering)
the configurations is around 4\% of the 
computation time in the smallest simulable lattice ($L=20$), and decreases 
steeply with size (it is less than 1\% of the total time for $L=30$).

Let us describe the main characteristics of a SUE board.
Devices used are listed in table \ref{comp},
apart from passive components (resistors, diodes, capacitors, leds, etc.).
The main electronic devices are the Altera 10K CPLDs~\cite{PLD}.

\begin{table}[b]
\small
\centering
\begin{tabular}{clll}
\hline\hline
Qty      & Type          &  Component    & Manufacturer \\ 
\hline
5        & CPLD          &  FLEX 10K30   & ALTERA       \\
1        & CPLD          &  FLEX 10K50   & ALTERA       \\
2        & PLD           &  EPM 7032-10  & ALTERA       \\
1        & PLD           &  EPM 7032-7   & ALTERA       \\
26       & SRAM          &  CY7C1031     & CYPRESS      \\
11       & LATCH         &  CY162841     & CYPRESS      \\
6        & PLL           &  CY2308-4     & CYPRESS      \\
1        & OSCILLATOR    &  SG615P       & SEIKO EPSON  \\
\hline\hline
\end{tabular}
\caption{Active Components in SUE}
\label{comp}
\normalsize
\end{table}

\begin{figure}[htb]
\centering
\vglue 0.3cm
\epsfig{figure=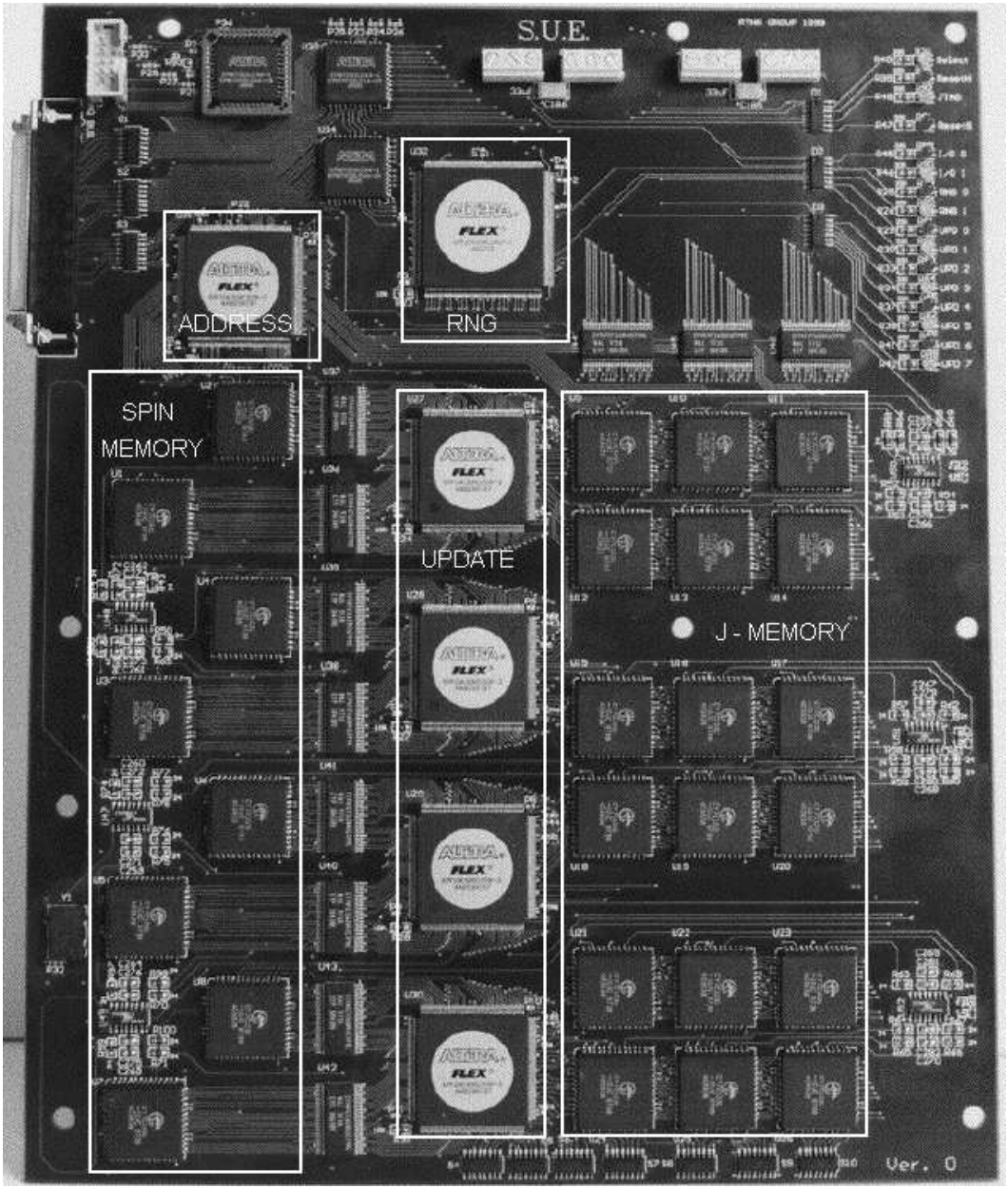,angle=0,width=140mm}
\caption{SUE board}
\label{suefoto}
\end{figure}

The photograph of one of the boards can be seen in fig.~\ref{suefoto}.
It contains four devices FLEX 10K30 responsible for the core of the Monte 
Carlo simulation ({\it UPDATE} area on the figure). That four
devices have the same electronic logic inside, which is prepared to
update two lattices in parallel.

On the right of these chips are the static memory devices (SRAM) which
store the couplings of the lattices ({\it J-MEMORY}). On the left, 
each {\it UPDATE} device 
has two SRAM devices which store the spin variables ({\it SPIN MEMORY}). 
Latch devices are 
used as tristate devices to manage the polarity of the data buses at high frequency. 

The {\it RNG} device is a FLEX 10K50 where a random number generator is 
programmed, allowing the use of canonic simulations.
Addressing of the memories and sincronization between the devices are
the main tasks of the fifth FLEX 10K30 ({\it ADDRESS}). The coupling memories
are addressed through latch devices to avoid fan-out problems.

External communication is provided by three EPM7032 chips placed 
near the 68-pin connector. One of them controls the board when the on-board 
programmable devices are not yet programmed. It responds to basic commands
sent from the HC, allowing to select and program the board.
The programmed logic establishes 4 control lines in each direction 
allowing communication between the HC an the {\it ADDRESS} device, and
a 32-bit data bus common to the HC and the {\it UPDATE} and {\it RNG} 
devices. The two lower bits in this bus reach {\it ADDRESS} too, and
act as extra control lines when needed. 

The clock signal is distributed to all the synchronous devices 
in the board through Cypress 2308-4 PLL devices.
In the upper right corner, a set of leds permits us to 
visualize the state of the machine. The connection to the HC is made 
through a 68 pin (SCSI-2 type) connector.
The SUE boards can share the same bus for an easy management from 
the HC.

Once the general architecture of a SUE board has been outlined,
the internal details are explained more deeply in the
next subsections.

\subsection{Updating Logic}

Four Altera 10K30 devices are responsible for the update. To each one 
of those devices lines are assigned to access the spin and 
coupling memories and the 32-bit bus through which the random 
number is provided.
That bus is used also to write and read the memories, the demon energy
or the probability table from the HC.

 In order to obtain
an updated spin every clock cycle we have designed a pipeline structure
that performs the algorithm step by step: A state machine runs over 
a 10 states cycle during the simulation, one spin being put into the 
updating pipeline at each step.

We have already mentioned that both algorithms are local. Indeed, the 
devices that actually perform the updating ignore where in the 
lattice is the site being updated, or the size of the system. They 
just process their input data and output the updated spins.
There is another device ({\it ADDRESS}) which takes care of the 
geometry and addresses the 
memories accordingly. That component is also responsible for stopping the
simulation when the desired number of configurations has been calculated. 

\subsection{Memory Scheme}

Two different memory banks, for spins and couplings, are available to each 
{\it UPDATE} device ({\it SPIN MEMORY} and {\it J-MEMORY} areas in 
fig.~\ref{suefoto}). 

Couplings are not dynamic variables (their 
values remain constant during the simulation), so the coupling memories 
are always in reading mode during update. When a site is to be updated, the 
couplings with its neigbours $x_+$, $x_-$, $y_+$, $y_-$, $z_+$, $z_-$ must be 
supplied to the updating engine. We organize therefore the 18 SRAM 
devices as a single bank of width $3\times 16$ bits and depth $6\times 64K$. 
Each lattice takes 6 of the 48 bits to store the six needed couplings. 
The maximum simulable volume is then limited by the depth to L=73.

The spins changing during the simulation, an appropiate mechanism is 
needed in order to read and write the configurations simultaneously: 
The spin memory is duplicated (P and Q banks), and while one memory bank is 
read from the other is written on. Because of that, two memory devices are connected 
to every {\it UPDATE} component, each one capable to store $64K \times$ 18 bits.

In order to understand how the spin memory is managed,
let us consider each column along the $x$ axis of the lattice divided 
in blocks of fixed length $l$.

To update one of the blocks, the block itself and its four $y$ and $z$
neighbours
have to be supplied. So, five blocks must be 
read to update one, implying that, if an updated spin every clock cycle is 
wanted, 
the block length has to be at least five (see subsection~\ref{pipeline}
below).

Each spin memory device of 18 bit words stores two lattices, so 9 bits 
are available for each lattice. Each block can contain from 5 to 9 spins, and 
the maximum lattice size $L\sim (l\times 64K)^{1/3}$ that can be stored is $68$, $73$
, $77$, $80$ 
or $84$, depending on the selected block size.
This limitation, together with the one we found from the coupling 
memory and the fact that the number of blocks must be even, yields 
the range of simulable sizes shown in table~\ref{SIZES}.

\begin{table}[h]
\small
\centering
\begin{tabular}{ccccc}
\hline\hline
l=5 & l=6 & l=7 & l=8 & l=9 \\ 
\hline
20 & 24 & 28 &    &    \\ 
30 &    &    & 32 &    \\ 
   & 36 &    &    & 36 \\ 
40 &    & 42 &    &    \\ 
   & 48 &    & 48 &    \\ 
50 &    &    &    & 54 \\ 
   &    & 56 &    &    \\ 
60 & 60 &    & 64 &    \\ 
   &    & 70 &    & 72 \\ 
   & 72 &    &    &    \\ 
\hline\hline
\end{tabular}
\caption{Simulable lattice sizes}
\label{SIZES}
\normalsize
\end{table}

The spin memories are arranged in the following way:
Each 9-bit word contains $l$ consecutive spins ($5{\le}l{\le}9$),
being consecutive (along x axis) lattice blocks stored in consecutive
memory addresses. The $V/l$ words are not read consecutively, 
but following a pattern that makes the block to be updated and its 
neighbouring blocks available to the {\it UPDATE} component, 
as explained in the next subsection.

On the other hand, the coupling memories store in 
the $n^{th}$ 6-bit word the couplings of
the $n^{th}$ spin with its six neighbours, and is read sequentially as
the $V$ spins are updated.

\begin{figure}[htb]
\centering
\vglue 0.3cm
\epsfig{figure=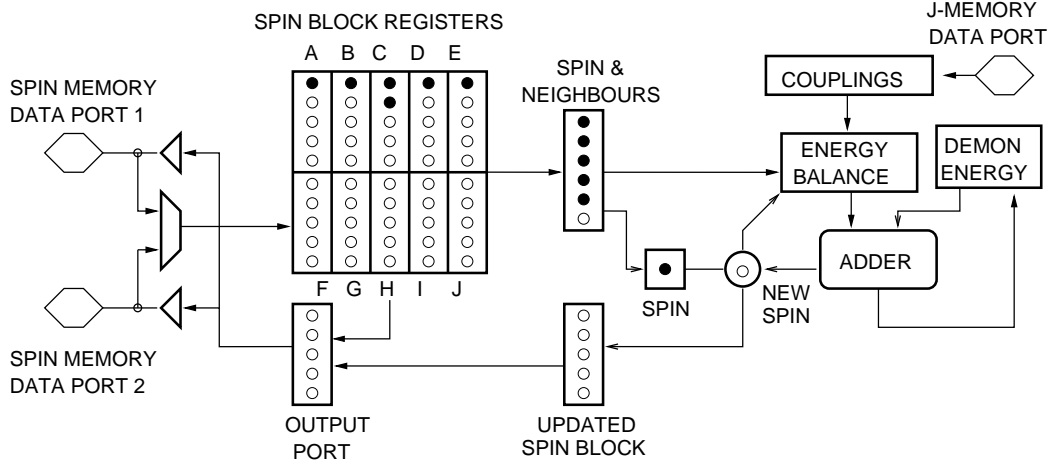,angle=0,width=140mm}
\caption{Demon algorithm pipeline implemented in the {\it UPDATE} devices}
\label{updatefig}
\end{figure}          

\subsection{Pipelined Updating \label{pipeline}}

In this subsection we describe the logic programmed in every {\it UPDATE}
device. We consider the case in which the demon algorithm is used
with a block length $l=5$ (see fig. \ref{updatefig}). In this case, 
the algorithm runs over a state machine with ten states. 
In each of those states a block is read from bank Q, which is in reading mode,
and stored in one of the internal registers A...J.
Let us suppose we have already been in states $0...4$, and some registers
are already loaded: A ($z_-$ neighbouring block), B ($y_-$ 
neighbouring block), 
C (block to be updated), D ($y_+$ neighbouring block) and E 
($z_+$ neighbouring block).

In state $5$, we send for update the first spin in the block stored in C.
The $x_+$ neighbour is in the same block, the $x_-$ neighbour is the
previous updated spin, which is still in the updating process
(and will not be needed until the last step), and the rest of the 
neighbours are stored in blocks A,B,D,E. 
We read simultaneously the $z_-$ neighbouring block of the next block to be 
updated and store it in register F.

In states $6$ to $8$, we continue sending for update the spins second to fourth
in block C, and loading registers G (next block $y_-$ neighbour), H (next 
block to be updated) and I (next block $y_+$ neighbour). 
In state $9$, the last spin in block C is sent into the update pipeline.
It is no longer true that the $x_+$ neighbour is in the same block, but it 
is in the block we have already stored in register H. Register J (next block 
$z_+$ neighbour) is loaded. When we return to state $0$, the updating of the 
block registered in H starts.

After some cycles, the updated value of the block that was stored in C
has been calculated and is written on the appropiate memory position in bank P,
in writing mode. 
Writing follows the same scheme as reading, not only the updated values 
are written but also the unchanged neighbours.

When a whole column (containing $L/l$ blocks of $l$ spins each) 
has been updated, we change the role of the memories:
we will now write on bank Q, the memory bank we were previously reading from, 
and read from bank P, the bank we were writing on. 
Bank P stores now the correct new configuration. Bank Q stores the old
configuration, which shall be overwritten with the result of updating the
column read from bank P.

We have seen that the writing and reading sequences are equal, although
writing is obviously delayed with respect to reading several cycles.
Due to this delay, to avoid problems in the change of 
role of the spin memory banks
the first block of the updated column, which should be read while its bank is
still in writing mode, is stored in a
{\it cache} memory inside the {\it UPDATE} devices. 
This mechanism requires at least four blocks, making the 
minimum simulable size to be $L=20$.

\subsection{Addressing Logic}

The {\it ADDRESS} device in fig.~\ref{suefoto} controls and 
addresses the memories, establishes the
functioning mode of the {\it UPDATE} and {\it RNG} devices and takes
charge of the communications with the HC. 
As we said above,
The board is accessed through a communication port with 32 bidirectional
lines devoted to data transfer and 8 control lines (4 in each direction).
Data lines are connected through tri-state circuits to the bus connecting
the {\it UPDATE} and {\it RNG} devices. Control lines are connected
to the {\it ADDRESS} chip, which controls the board according to the 
commands sent from the HC.
 
The implemented instruction set allows us to:
\begin{itemize}
\item{}program the devices.
\item{}read/write the spin and coupling configurations.
\item{}read/write the demon energies.
\item{}load the number generator initialization table. 
\item{}load the probability tables used in the Heat Bath algorithm.
\item{}set the number of iterations to run.
\item{}start the simulation.
\end{itemize}

The {\it ADDRESS} device controls the {\it UPDATE} and {\it RNG} 
devices to carry out
that operations. A 3-bit wide bus is used to encode the instructions
for the {\it UPDATE} devices.

\subsection{Random Number Generator}

The Altera 10K50 device ({\it RNG} chip in fig.~\ref{suefoto}) 
is a 32-bit pseudo-random number generator of the 
R250 kind. Those generators are known to suffer some problems in 
Monte Carlo simulations, but only with non-local algorithms~\cite{RNG}.

In the C implementation, a vector is initialized with a conventional 
pseudo-random number generator. Using the macro instruction RANDOM we 
run over the wheel, getting a new number and changing one of the values 
in the wheel:

\begin{verbatim}
#define RANDOM  ( (irr[ip++]=irr[ip1++]+irr[ip2++])^irr[ip3++] )
\end{verbatim}

The variables involved need to be properly initialized before using the
defined macro:

\begin{verbatim}
/* random number generator initialization */

unsigned int irr[256];
unsigned char ip, ip1, ip2, ip3;

ip=128;
ip1=ip-24;
ip2=ip-55;
ip3=ip-61;

for(i=0; i<256; i++)
    irr[i]=(unsigned int) rand();    
\end{verbatim}

\begin{figure}[htb]
\centering
\vglue 0.3cm
\epsfig{figure=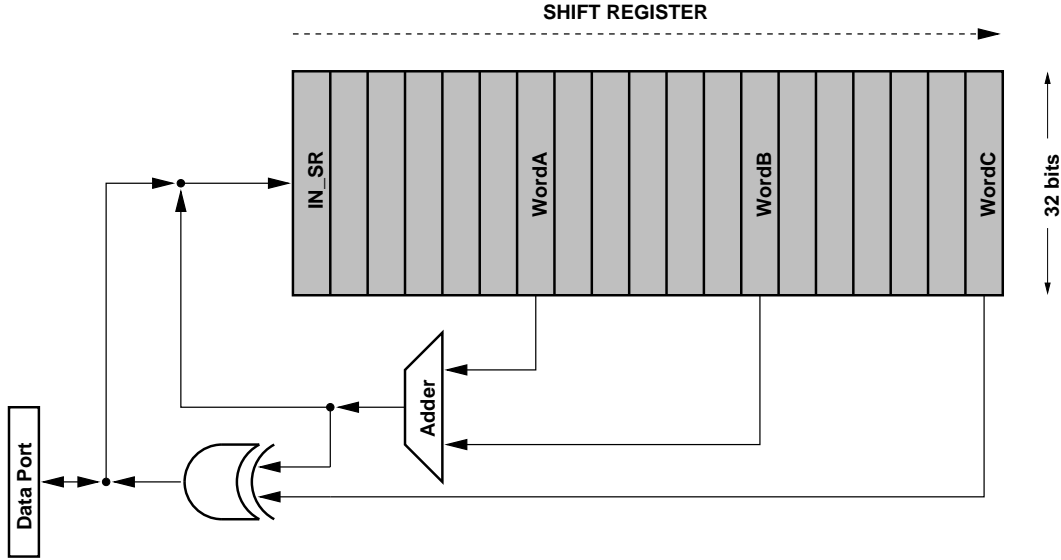,angle=0,width=140mm}
\caption{Random Number Generator}
\label{rngfig}
\end{figure}          

In the {\it RNG} device (see Fig.~\ref{rngfig}), 
the \verb+irr[i]+ wheel becomes a 32-bit wide shift 
register, reproducing that way the effect of incrementing 
{\tt ip},{\tt ip1},{\tt ip2} and {\tt ip3}. 
An adder sums the words {\tt WordA} and {\tt WordB} and stores the 
result in the first position {\tt IN\_SR}. This result also 
serves as input to a XOR function, together 
with the value in the last register {\tt WordC}. The result of this 
function provides us with the pseudo-random number, every clock cycle.

The seeds loading process is controlled by the {\it ADDRESS} component,
which also enables the random number generation during the simulation.

\subsection{Software}

The boards are connected to the HC through a data acquisition card 
PCI-DIO32HS from National Instruments. To access the DAQ, a Linux driver 
has been programmed, and also a user library allowing to operate with the
boards in an easy way.

The functions available to the user are the following:

\begin{itemize}
\item{}{\tt dioinit}  :  initializes the DAQ boards to be used by the HC.
\item{}{\tt boardsel} :  selects one board among those connected to the HC.
\item{}{\tt ws}       :  writes the spin configuration corresponding to one of
		the {\it UPDATE} devices in the selected board.
\item{}{\tt rs}       :  reads the spin configuration.
\item{}{\tt wj}       :  writes the couplings of the lattices in the selected board.
\item{}{\tt rj}       :  reads the couplings in the selected board.
\item{}{\tt rd}       :  reads the Demon energy.
\item{}{\tt wd}       :  writes the Demon energy.
\item{}{\tt wmesfr}   :  sets the number of iterations in each run.
\item{}{\tt wrng}     :  writes the initial random number table. 
\item{}{\tt wprob}    :  writes the probability table \ref{partition} on {\it UPDATE}.
\item{}{\tt startsue} :  starts the simulation in the selected board.
\item{}{\tt waitsue}  :  waits for one of the boards to finish.
\end{itemize}

The functions to access the memories get the arguments as arrays of bytes, 
where each element corresponds to one site in the 
lattice and the eight bits in the element to each one of the lattices in 
the board, as is usual in multi-spin code. The user needs not worry about
SUE internal details.

\subsection{Design considerations}

\subsubsection{Programming method}

CPLDs are electronic devices which can be programmed as many
times as needed. They lose their program code every time the board
is switched off, so they have to be reprogrammed after switch on.
To manage the programming task, these devices are connected sequentially
creating a JTAG chain which is controlled by the HC through the 
communication port described above. No extra cables are needed,
providing easy on-board reprogrammability controlled from the HC.
This feature was extremelly important during the debug process
of the boards

\subsubsection{Printed Circuit Board}

The printed circuit board surface is $24.5 \times 30.5 cm^2$, 
and it is $2mm$ thick.
Manufactured in FR4 fiber, it consists of eight layers (four dedicated to 
signal transimision and the others to powering).

The board satisfies the ATX standard. In the final version, the 12 boards 
are mounted in a rack and are fed by a 800 Wat source at 5 V. Current, 
voltage and temperature are monitored. Full operation values are 90 Amp at 5 V.

\subsubsection{Frequency}

The proper working of the circuit requires perfect synchronization 
between the active devices. The working frequency is 48 MHz, and the 
clock signal should reach the 32 components spread over a $747cm^2$ surface.

The clock distribution is made through CY2308-4 devices (3.3V Zero 
Delay Buffers), provided with a PLL mechanism ({\it Phase Locked Loop})
that allows to double the input frequency and supply eight outputs.

A 12~MHz oscillator is connected to a PLL device that doubles its frequency. 
Five outputs are driven into PLL components that double the frequency again and
feed the neighbouring components. In this way, the clock is distributed 
across the circuit at low frequency, and the frequency doubled near the 
final components.

\subsubsection{Transmission lines}

As a consequence of the large size of the circuit, there exist connections
with a large total trace length. The rise times of the signals determine 
whether the transmission line behaves like a distributed circuit or not.

The effective length associated with the rise time of a signal is 
\begin{equation}
l=\frac{T_r}{D}
\label{rise}
\end{equation}
where $T_r$ is the rise time and $D$ the propagation delay, characteristic 
of the material. We must consider a distributed circuit if the length of
the transmission line is greater than a quarter of the effective length.

In our board, diode barriers protect the traces addressing the
coupling memories, the data bus connecting {\it UPDATE} and {\it RNG} 
devices, and the connector for external communication.
The rest of the signals, generated by memory devices or Altera 10K components
(the latter allow the user to set the rise time), have rise times short
enough for the system to behave in a lumped fashion.  

\section{Development Process \label{DEVELOPMENT}}

Initially, only one board was manufactured. In a first stage, we tested its 
general performance. After being able to communicate with the machine, the 
programming mechanism was implemented. Different test programs were written 
and compiled to program the CPLDs, using Altera's MAXPlus+II development 
enviroment. Once we checked that all the components worked properly (fixing
some electrical bugs in the way), the Demon algorithm was progressively 
implemented. We chose the demon algorithm because it is microcanonical, so 
the random number generator is not needed and the conservation of the total
energy provides a fast test mechanism. Additional functionalities were added
and the algortithm scheme was fine-tuned, until the program was complete.

When the rest of the boards were available, they were tested with this Demon
program, and the Heat Bath algorithm was then implemented.
The structure of the algorithm remained almost the same, although some details 
in the algorithm had to be changed and some new functions were added to the 
user library. The main novelty was the random number generator usage, which 
worked finally with an appropriate pipelining scheme both in RNG and UPDATE
chips.

In the early debugging stage, development had been carried at 24~MHz, so we 
switched 
to high frequency. Some fine-tuning in the programs was needed and the CPLD 
logic layout was carefully studied in order to reach the design-goal of 48~MHz.

To make sure of the proper working of the machine beyond 
any doubt, an emulator 
was developed to run in a PC, so the machine configurations can be compared 
with those obtained with the PC emulation. This test proved that both the
updating algorithm and the random number generator worked as intended.

\section{Performance}

Table~\ref{PERFORM} compares the update speed achieved by SUE 
with that of some simulations run by our group in different computers, 
and with the performance obtained running highly optimized 
multi-spin code in a Cray T3E supercomputer as reported in~\cite{CARA}.
We can see that the whole machine matches the computational power of 
one hundred processors of a CrayT3E.

\begin{table}[h]
\small
\centering
\begin{tabular}{lr}
\hline\hline
System & Update speed (ns/spin)\\ 
\hline
Pentium Pro 200 MHz 	&	170 \\
Pentium II  500 MHz 	&	102 \\
Alpha 133 MHz       	&	215 \\
Alpha 400 MHz       	& 	 58 \\
Alpha 500 MHz       	&	 44 \\
APE (tower)         	&	  6 \\
Alpha EV5, 600 Mhz \cite{CARA}  &	22 \\
SUE (single board)  	&	2.6 \\
SUE (twelve boards)	&	0.22 \\
\hline\hline
\end{tabular}
\caption{SUE performance}
\label{PERFORM}
\normalsize
\end{table}

\section{Preliminary Physical Results}

In this section we present some preliminary results obtained with SUE.
We have run an $L=20$ lattice in $4$ boards at $48$ Mhz. We have 
simulated $1600$ sets of $\{J_{i,j}\}$, with $2$ {\it replicas}
at $12$ different values of $\beta$, which previously have been controlled
to have a correct transfer probability between them with the parallel
tempering method. We measure every $16384$ sweeps, collecting $800$
measurements. These results have been obtained in $60$ days.

\begin{figure}[t]
\vglue 0.3cm
\epsfig{figure=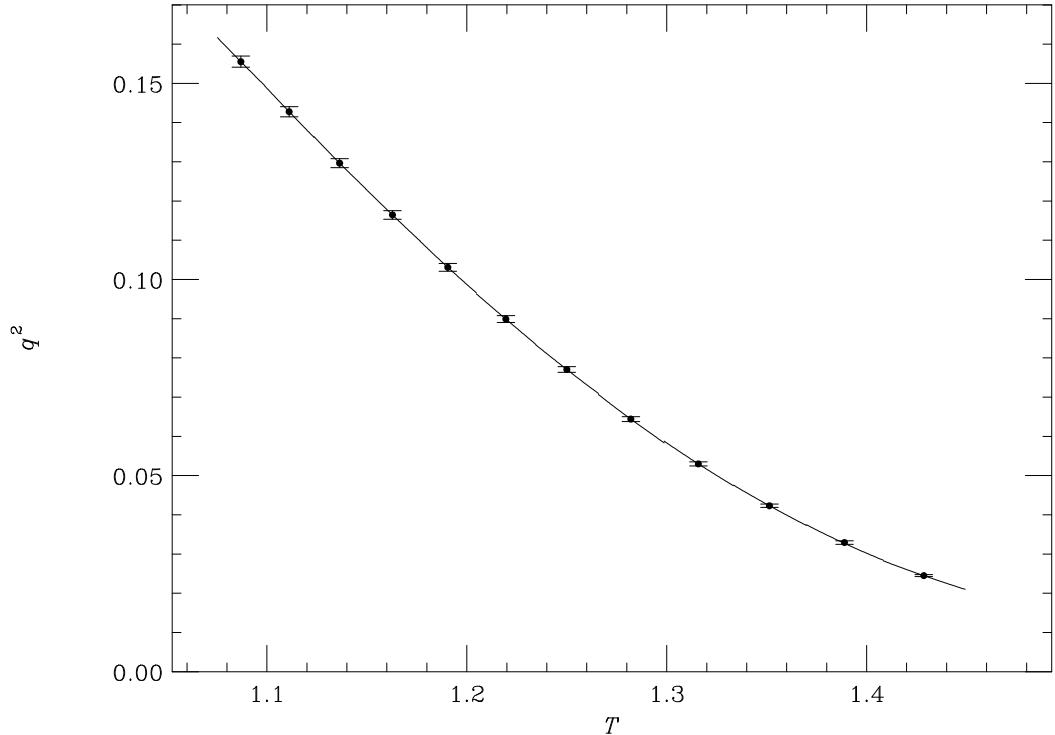,angle=90,width=140mm}
\caption{Overlap in $L=20$, as a function of $T$. The points correspond to
the $12$ simulated $T$, and the lines are obtained from the spectral density
method.}
\label{phis}
\end{figure}

In Fig. \ref{phis} we plot the value of the average squared
overlap.

The errors are plotted only at the simulations points. We are working around
the critical region, as we reach high values for $q^2$. The extrapolated lines
connect properly and the different values evolve smoothly for different
$T$ values, as corresponds to a good thermalization and a high 
transition probability from parallel tempering.

At the moment of writing, we are running $L=20$ in 12 boards and almost finished the 
runs. Afterwards we will start the simulation in the $L=30$ system, and estimate that
the time needed to obtain good results is around one year.

{\bf Acknowledgements}

We wish to thank H.G.~Ballesteros, J.M.~Carmona, L.A.~Fern\'andez, D.~I\~niguez, 
and J.J.~Ruiz-Lorenzo for useful discussions.
Partially supported by DGA (P46/97) and CICyT (AEN97-1768 and AEN99-0990).


\begin{thebibliography}{9}

\bibitem{APE} 
The APE Collaboration, 
{\em Comp. Phys. Com.} {\bf 57} (1989) 285. 

\bibitem{CHRIST} 
N.~H.~Christ and A.~E.~Terrano,
{\em IEEE Trans. Comput.} {\bf 33} (1984) 344. 

\bibitem{RTN} 
The RTN Collaboration,
{\em Procc. of CHEP 92} CERN 92-07. 

\bibitem{HOOGLAND} 
A.~Hoogland, J.~Spaa, B.~Selman and A.~Compagner,
{\em J. Comp. Phys.} {\bf 51} (1983) 250. 

\bibitem{SUE2D} 
J.~Pech, A.~Taranc\'on and C.L.~Ullod, 
{\em Comp. Phys. Com.} {\bf 106} (1997) 10, {\em hep-lat/9611014}. 

\bibitem{SG}    
M.~Mezard, G.~Parisi and M.~A.~Virasoro,
{\em Spin Glass Theory and Beyond}
(World Scientific, Singapore 1987).

\bibitem{TEMP}
E.~Marinari, G.~Parisi and J.J.~Ruiz-Lorenzo, 
in {\em Spin Glasses and Random Fields} 
(World Scientific, Singapore 1998), {\em cond-mat/9701016}.

\bibitem{MARI}
E.~Marinari, G.~Parisi and J.J.~Ruiz-Lorenzo, 
{\em Phys. Rev. B} {\bf 58} (1998) 14852, {\em cond-mat/9802211}.

\bibitem{BERG}
B.A.~Berg and W.~Janke,
{\em Phys. Rev. Lett.} {\bf 80} (1998) 4771.

\bibitem{YOUNG}
N. Kawashima and A. P. Young, 
{\em Phys. Rev. B} {\bf 53} (1996) R484, {\em cond-mat/9510009}.

\bibitem{FISCHER}       
M.~E.~Fischer and A.~Nihat Baker,
{\em Phys. Rev. B} {\bf 26} (1982) 2507. 

\bibitem{QFC} 
M.~Creutz, 
{\em Quantum Fields on the Computer}
(World Scientific, Singapore 1992).

\bibitem{CREUTZ} 
M.~Creutz, 
Microcanonical Monte Carlo Simulation.
{\em Phys. Rev. Lett.} {\bf 50-19} (1993). 

\bibitem{CluJJ}
J.J.~Ruiz-Lorenzo and C.L.~Ullod,
{\em Comp. Phys. Com.} {\bf(125) 1-3} (2000) 210, {\em cond-mat/9812378}.

\bibitem{PLD} 
Altera Corporation,
{\em Altera Data Book} 1995. 

\bibitem{RNG}
H.G.~Ballesteros and V.~Martin-Mayor, 
{\em Phys. Rev. E} {\bf 58} (1998) 6787, {\em cond-mat/9806059}.

\bibitem{CARA}
M-~Palassini, S.~Caracciolo,
{\em Phys. Rev. Lett.} {\bf 82} (1999) 5128, {\em cond-mat/9911449}.

\end{thebibliography}
\end{document}